\documentclass[12pt,noinfoline]{imsart}

\usepackage{fancyhdr, geometry}
\usepackage{booktabs}
\usepackage{url}
\usepackage{graphicx}
\usepackage{comment}
\usepackage[normalem]{ulem}
\usepackage{amsmath,amsfonts,amssymb,amsthm}
\usepackage{xcolor}
\usepackage{mathtools}
\usepackage{hyperref}
\usepackage{bbm}
\usepackage{pgfmath}
\usepackage{tikz}
\usepackage{caption}
\usepackage{subcaption}
\usetikzlibrary{backgrounds}
\usetikzlibrary{backgrounds, intersections, calc}

\usepackage{algorithm}

\usepackage{algpseudocode}

\newcommand{\fiber}[1]{\mathcal{F}_{#1}} 

\makeatletter
\renewcommand{\@fnsymbol}[1]{\ifcase#1\or *\or \dagger\or \ddagger\or
   \mathsection\or \mathparagraph\or \|\or **\or \dagger\dagger\or \ddagger\ddagger \fi}
\makeatother

\newtheorem{lemma}{Lemma}[section]

\newtheorem{definition}{Definition}[section]

\makeatletter
\def\@fnsymbol#1{\ensuremath{
  \ifcase#1\or *\or \dagger\or \ddagger\or
  \mathsection\or \mathparagraph\or \|\or **\or \dagger\dagger\or \ddagger\ddagger
  \else\@ctrerr\fi}}
\makeatother

\begin{document}


\begin{frontmatter}

\title{Multilevel Sampling in Algebraic Statistics}

\author[A]{Nathan Kirk}
\author[A]{Ivan Gvozdanovi\'c}
\author[A]{Sonja Petrovi\'c\thanks{Corresponding Author: \texttt{sonja.petrovic@illinoistech.edu}}}

\address[A]{Department of Applied Mathematics, Illinois Institute of Technology, Chicago IL, USA}

\begin{abstract}
This paper proposes a multilevel sampling algorithm for fiber sampling problems in algebraic statistics, inspired by Henry Wynn’s suggestion to adapt multilevel Monte Carlo (MLMC) ideas to discrete models. Focusing on log-linear models, we sample from high-dimensional lattice fibers defined by algebraic constraints. Building on Markov basis methods and results from Diaconis and Sturmfels, our algorithm uses variable step sizes to accelerate exploration and reduce the need for long burn-in. We introduce a novel Fiber Coverage Score (FCS) based on Voronoi partitioning to assess sample quality, and highlight the utility of the Maximum Mean Discrepancy (MMD) quality metric. Simulations on benchmark fibers show that multilevel sampling outperforms naive MCMC approaches. Our results demonstrate that multilevel methods, when properly applied, provide practical benefits for discrete sampling in algebraic statistics.
\end{abstract}

\end{frontmatter}

\maketitle

\maketitle

\section{Introduction}

Henry Wynn's work in bringing modern algebraic geometry to the field of statistics laid the foundation for the development of algebraic statistics as a subject. The term `algebraic statistics' was coined in his joint work  \cite{PistoneWynnEva} (see also the monograph \cite{WynnAlgStatBook2001}) and broadly denotes the application of nonlinear algebraic and nonconvex geometric methods to statistical problems. 
Although algebraic and geometric ideas had long been present in statistics, the predominant tools before the 2000s were linear algebra and convex geometry. A shift began in the mid-1990s: Pistone and Wynn \cite{PistoneWynn} employed Gröbner bases to study confounding in experimental design, while Diaconis and Sturmfels \cite{DS98} used them for performing exact conditional tests. Wynn and collaborators went further, recasting core concepts in statistics--such as optimal design and system reliability theory \cite{reliability}--within the framework of computational algebra.
Today, Wynn's influence remains central in these areas. A central theme of his work was bridging interdisciplinary language barriers to connect ideas across fields.
This paper presents one such effort\footnote{In 2016, Henry Wynn visited Chicago twice. During discussions with S. Petrovi\'{c}, F. Hickernell, and their students, he proposed a new approach to sampling lattice points in polytopes, inspired by multilevel methods.}: the adaptation of multilevel sampling \cite{Gil15a}--a powerful tool in Monte Carlo methods employed to reduce computational cost when approximating expectations of quantities of interest, most commonly applied to path simulation \cite{Gil08a}--to the setting of sampling discrete structures, a core concern in algebraic statistics.

The discrete structure Wynn was interested in sampling is an integer lattice in a polytope:
\begin{equation}\label{eq: lattice in polytope}
    \fiber{A}(b):=\{x\in\mathbb Z_{\geq 0}^n:  Ax=b,  A\in\mathbb Z^{d\times n},b\in\mathbb Z^d\}. 
\end{equation}
The statistical motivation for sampling polytopes is a cornerstone of algebraic statistics, and the statistical ideas for conditional inference that crucially rely on this problem were developed by Ronald Fisher; see Section~\ref{sec:background on fibers}. 

Sampling algorithms and strategies for exploring the set \eqref{eq: lattice in polytope} abound. 
These include, for example, the hit-and-run sampler \cite{hit-and-run} (see also \cite{DiaconisAnderson}), sequential importance sampling (SIS, see \cite{Liu2001reprint} as a standard reference), 
and a hybrid approach that combines SIS and Markov bases \cite{KahleYoshidaGarciaPuente}. 
\cite{Chen_etal-lattice-contingency} reviews and compares a Markov chain implementation and SIS for logistic regression models, showing how SIS outperforms in certain scenarios; while \cite{Dob2012} demonstrates the superiority of Markov bases over SIS in general fibers. 
Dobra also focused on the problem of actually computing Markov bases in practice, which is intractable for many problems of reasonable size that arise in applications. Indeed, there have been many attempts to work without the entire basis;  see \cite{HAT} for an early idea based on remarks made in \cite{DS98}. The fact of the matter is that using a proper subset of a Markov basis necessarily disconnects \emph{some} fibers, and it is impossible to tell in general which ones. A recent approach to circumvent the problem is a Bayesian framework, called RUMBA \cite{RUMBA}, that learns combinations of smaller basis elements. Given recent advances in reinforcement learning, another reasonable way to try to solve the problem is by actor-critic algorithms; this has been proposed in \cite{RLFiberSampler}, motivated by \cite{DeLoeraWu-RL}. 
Today, the fiber sampling problem still faces some open challenges: training agents comes at a cost, tuning sampler parameters takes time, and mixing of Markov chains is difficult to evaluate in a timely manner. 

In this paper, we pay homage to Wynn’s idea of introducing multilevel sampling to the specific problem of sampling lattices of relevance in algebraic statistics (Section~\ref{sec:background on fibers}). The new sampling procedure, introduced in Section~\ref{sec:algo}, is accompanied by a metric for evaluating sample quality, detailed in Section~\ref{sec:measuring sample quality}. This eliminates the need for traditional convergence analysis, as is typically required in Markov chain-based methods. The method is tested on canonical examples in algebraic statistics that are known to be challenging due to sparsity: although the observed data lie in low dimension (e.g., $4 \times 4$ contingency tables), the fibers contain on the order of 185 million lattice points. Simulation results are presented in Section~\ref{sec:simulations}. We hope that this work encourages broader adoption of multilevel procedures in algebraic statistics.

\subsection{Problem and notation}\label{sec:background on fibers}


The problem we solve is the following: 
\emph{Obtain a random sample from the fiber defined in Equation~\eqref{eq: lattice in polytope} so as to be able to estimate the sampling distribution of a statistic.}

Most of our readers will be familiar with the fact that conditional distributions given sufficient statistics in log-affine models, as defined in \cite{Lauritzen}, are supported exactly on lattice points in polytopes. 
Log-affine models are exponential family models (\cite{Barndorff-Nielsen}) for $k$ discrete random variables $X_1,\dots,X_k$ with finite state spaces, $X_i\subset[d_i]:=\{1,\dots,d_i\}$. An exponential family being log-affine implies the existence of a large, often sparse matrix $A=[a_1\dots a_n]\subset \mathbb Z^{d\times n}$, which computes the model's sufficient statistics:  the family  of probability distributions $\mathcal M_A:=\{p_{\theta}\}_{\theta\in \Theta}$ takes the form:  
	\[
		p_i = \frac{1}{\psi(\theta)}\exp\big(\langle\theta,a_i\rangle\big), 
	\]
with parameter space $\Theta\subset\mathbb R^n$, and partition function $\psi(\theta)$. 
We may assume that $1\in rowspan(A)$, a condition that can be fixed by rescaling model parameters. In algebraic statistics, the model $\mathcal M_A$ is called a \emph{log-linear} model.

One canonical application for constructing a sample from \eqref{eq: lattice in polytope} is to perform the test of goodness-of-fit for $\mathcal M_A$ when asymptotic considerations fail--namely, in high-dimensional or large but possibly sparse state spaces. 
Early references such as \cite{haberman1976generalized, Habe:1981, Haberman77, Haberman88} include discussions inadequacy of standard asymptotic approximations are  for estimating the model fit $p$-value based on the approximate distribution of the chi-squared statistic. 
Thus, in our demonstrations, we focus obtaining the sampling distribution of the chi-square statistic, as it can be used as a goodness-of-fit statistic for testing the fit of the model to the observed data. The distribution on the fiber can be either uniform (for conditional volume tests \cite{DiaconisEfron-volumeTest}) or the conditional distribution given sufficient statistics (uniform or often hypergeometric, depending on the table cell sampling scheme \cite{FienberBookCategorical}). 
Measuring distance of observed data to the set of joint observations consistent with the log-linear model naturally lead to conditioning on sufficient statistics,  removing dependence on model parameters; conditional inference was described by Agresti in 1990 (see \cite{Agresti2002,Agresti}). 
This was the context considered in \cite{DS98}, who introduced Markov bases for solving this broad class of sampling problems. 
The crucial fact is that a Markov chain is built using moves that have analogues in polynomial algebra and polyhedral geometry. Each instance inherits a finite set of moves necessary to obtain an irreducible chain for \emph{any} value of the sufficient statistic. Diaconis and Sturmfels named any such finite collection of moves a \emph{Markov basis}; see \cite[Theorem 3.1]{DS98}. Their seminal paper has over 1000  citations; we point the readers to the comprehensive references \cite{AHT2012} and \cite{SethBook} and a recent survey \cite{MB25years}. 

To fix the notation, we will refer to the set ${\fiber{A}(b)}$ from Equation~\eqref{eq: lattice in polytope} as the $b$-fiber of $A$. Elements  
${\mathcal{B}}\subset \ker_{\mathbb{Z}}(A)$, which by definition preserve the value of the sufficient statistics, are called \emph{moves} on the fiber.  
For given moves $\mathcal B$, one can construct a fiber graph with vertices points in the fiber and edges $(u,v)$ 
for all $u-v\in \pm\mathcal{B}$. 
Given a fixed $A$, a set $\mathcal B$ is called a \emph{Markov basis of $A$}  if this fiber graph is  connected for \emph{all} vectors $b$. 
That such a set exists and is finite is a consequence of the Hilbert Basis Theorem, invoking the  Fundamental Theorem of Markov bases \cite{DS98}. 

Any  proper subset of a Markov basis is likely to leave at least one fiber disconnected for a given $A$. The easiest subset to compute, which is at least needed to generate the lattice of correct dimension, is the \emph{lattice basis of $A$}, which is simply the linear-algebra basis of the vector space $\ker_{\mathbb Z}(A)$. It is known that lattice bases leave  fibers disconnected in general.  On the other hand, work as early as \cite{Chen_etal-lattice-contingency} proves that lattice bases suffice for sampling a \emph{relaxation} of the fiber in the case of logistic regression models; this idea was also successfully explored for no-3-factor models for small tables by \cite{Yoshida-Barnhill:Negative-cell}. However, the negative results in \cite{MB25years} warn against the generalization of this idea: one has to choose the cells allowing to go negative very carefully, otherwise the chain remains disconnected for arbitrary relaxations.

\section{Multilevel Monte Carlo Methods}
\label{sec:background on multilevel}

Monte Carlo methods estimate quantities of interest by averaging random samples. When the target quantity can be expressed as an expectation, such as $\mathbb{E}[g(\mathbf{X})]$ with $\mathbf{X} \sim \pi$, a simple Monte Carlo estimator takes the form $\frac{1}{n} \sum_{i=1}^{n} g(\mathbf{X}_i)$,
using $n$ independent samples $\mathbf{X}_i \sim \pi$. These estimators are unbiased and converge at a rate of $\mathcal{O}(n^{-1/2})$, regardless of dimension. However, the computational cost is typically $\mathcal{O}(dn)$, which becomes prohibitive in high-dimensional problems, for example when $d$ corresponds to a fine discretization in time or space.

Multilevel Monte Carlo (MLMC) methods \cite{Gil15a} reduce this cost by using a hierarchy of approximations to the quantity of interest. Rather than estimating $\mathbb{E}[g(\mathbf{X})]$ directly at high resolution, MLMC expresses it as a telescoping sum over a sequence of increasingly accurate (and expensive) models:
\[
\mathbb{E}[Y_L] = \mathbb{E}[Y_1] + \sum_{l=2}^L \mathbb{E}[Y_l - Y_{l-1}],
\]
where each $Y_l$ approximates $g(\mathbf{X})$ at level $l$. Crucially, since the differences $Y_l - Y_{l-1}$ have lower variance than $Y_L$, fewer samples are needed at the more expensive levels. Thus, MLMC reduces the total cost to that of solving a low-fidelity approximation.

This cost saving regime has therefore been applied to many other domains, e.g., \cite{HerSch20a, Vro19a} despite originally developed for estimating expectations in numerical simulation (e.g., for path dependent finance problems and PDEs). MLMC has more recently been extended to the approximation of distribution functions and densities in \cite{gil15_multi_density}. These MLMC generalizations rely on smoothing techniques more sophisticated than standard kernel density estimation. Estimating distribution functions or densities from multilevel procedures directly involves discontinuities or singularities (from e.g., subtraction of indicator functions or Dirac deltas respectively), which can lead to high variance between levels and bias. By replacing these with smooth approximations, MLMC maintains variance decay and ensures stable convergence; see Section \ref{sec:algo} and \ref{sec:theoretical} for full details.

In this paper, we consider an application introduced to us by Henry Wynn: how to reinterpret multilevel sampling in the context of discrete structures, where the objects of interest are no longer continuous functionals but integer-valued points constrained by combinatorial or geometric conditions.  Importantly, the benefit of this multilevel perspective does not lie in reducing the cost of high-resolution discretization, as in traditional applications, but rather in improving the coverage and exploration of the lattice space. By structuring the sampling hierarchy across increasingly refined subsets of the lattice, we aim to leverage multilevel sampling to better navigate complex discrete spaces.

\section{A Multilevel Algorithm for Fiber Sampling}\label{sec:previous_multilevel}
Traditionally, the most used method for sampling fibers has been Metropolis-Hastings (MH) algorithm and its variations. With the introduction of Markov basis \cite{DS98}, such class of algorithms seemed like an obvious and cheap way to obtain samples on the high dimensional lattice bounded by a polytope $\mathcal{P}$. Nevertheless, these algorithms operate on a single sampling level, and their goal is not to efficiently explore the fiber but simply to obtain a sample which follows the proposed distribution. The closest attempt at multilevel fiber sampling was proposed as a lemma in \cite{DS98}. 

\begin{lemma}[Lemma 2.2, \cite{DS98}]\label{lemma: diaconis_multistep}
Let $\{f_1, \dots, f_L\}$ be a Markov basis for a model $\mathcal{M}_A$, let $t$ be the observed sufficient statistic. 
$\fiber{A}(t)$ denotes the fiber of tables with sufficient statistic $t$. Generate a Markov chain on $\fiber{A}(t)$ by choosing $I$ uniformly in $\{1, \dots, L\}$. If the chain is currently at $g \in \fiber{A}(t)$, determine the set of $j \in \mathbb{Z}$ such that $g + j f_I \in \fiber{A}(t)$. Choose $j$ in this set with probability proportional to
\[
\prod_{x \in C_I} \frac{1}{[g(x) + j f_I(x)]!},
\]
where $C_I = \{x : f_I(x) \neq 0\}$. Then, this is a connected, reversible, aperiodic Markov chain on $\fiber{A}(t)$ with the correct (i.e., hypergeometric) stationary distribution. \end{lemma}

The sampling procedure described in Lemma \ref{lemma: diaconis_multistep} defines a valid Markov chain Monte Carlo method, albeit one with acceptance probability $1$. By sampling the step size $j$ from a weighted distribution tailored to the target (hypergeometric) distribution, the algorithm avoids explicit accept/reject steps while preserving reversibility and ensuring the correct stationary distribution. The authors further illustrate how a different stationary distribution on the fiber can be reached by this Markov chain by changing the probability of choosing the $j$s.
Despite these theoretical guarantees, this sampling procedure like other MH-type methods still requires a burn-in period and long chain lengths to ensure reliable convergence. 

\begin{figure}[t] 
\centering
 \includegraphics[width=0.45\linewidth]{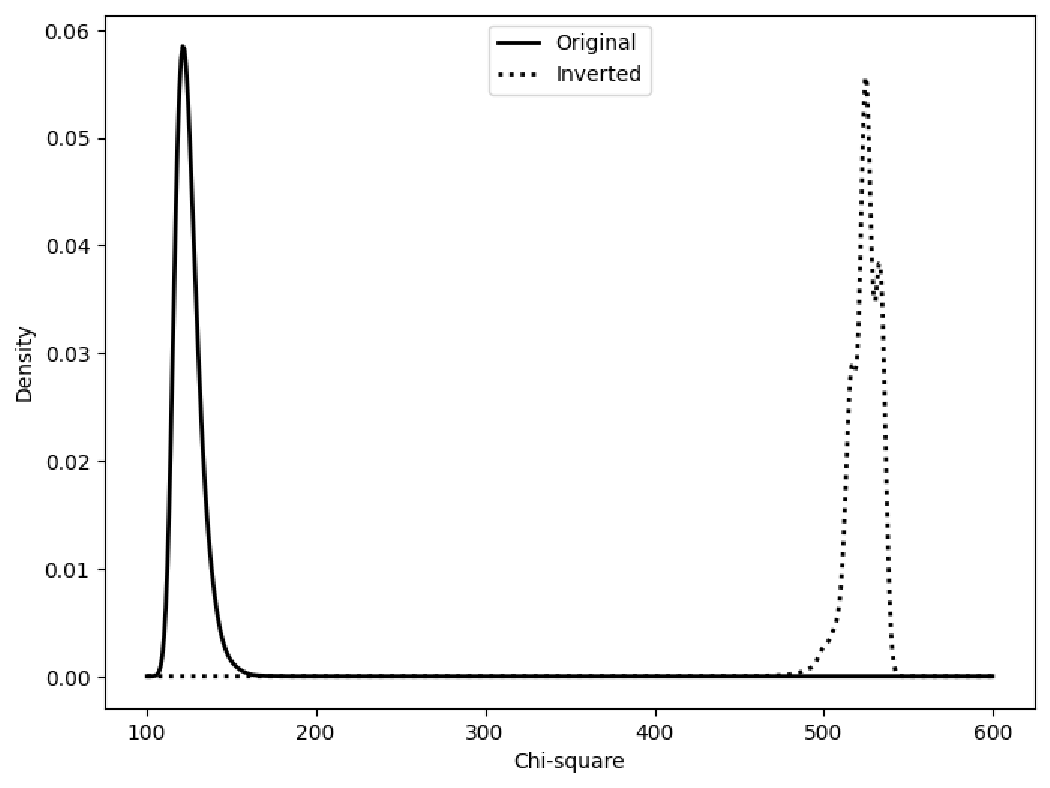}   
\caption{Smoothed empirical distributions of the chi-square statistic using the original sampler from Lemma \ref{lemma: diaconis_multistep} and the inverted step-size distribution.
Both experiments use the same fiber described in the Appendix \ref{app:fiber} with $n=100,000$. Estimates are smoothed with a Gaussian kernel.}
\label{fig: diaconis_multistep_fig} 
\end{figure}

One may be tempted to use this result as a basis for a multilevel sampler. However, this method is more like a Gibbs sampler (or hit-and-run), rather than a multilevel sampler. In fact, attempting to control the sizes of moves in the Markov chain by changing the multiples $j$ for any particular Markov move using this result looses control of the stationary distribution. To illustrate this issue (see Appendix \ref{app:fiber} for details of the fiber used), we compare the smoothed empirical distributions of chi-squared statistics obtained from two samplers: 1) the original sampler described in Lemma \ref{lemma: diaconis_multistep} and 2) a modified version in which the step size $j\in\mathbb{Z}$ is chosen with probability proportional to
\[
\prod_{x \in C_I} [g(x) + j f_I(x)]!,
\]
over the set of feasible \( j \). The results are shown in Figure \ref{fig: diaconis_multistep_fig}.

The original sampler's stationary distribution of the chi-square statistic is induced by the hypergeometric distribution of the fiber elements, as per Lemma~\ref{lemma: diaconis_multistep}. Notice that smaller step sizes $j$ are favored, which causes the chain to linger near its starting point. The second sampler prioritizes larger step sizes $j$ by inverting the probabilities; however, the stationary distribution is, by design, induced by another distribution on the fiber. These observations underscore a central limitation: \textit{without careful tuning of step sizes, naive MH samplers may fail to produce representative samples of the fiber, or may converge to a different stationary distribution.}

To address this, we now propose a multilevel sampling algorithm, adapted to estimate densities from samples generated over a discrete space, that avoids unnecessary burn-in while still producing samples informative for tasks such as goodness-of-fit testing. Our approach follows \cite{gil15_multi_density} and combines ideas from both Metropolis–Hastings and multilevel Monte Carlo (MLMC) methods. We focus on fibers which, although a relative term, could be called dense and use moves coming from either Markov basis or lattice basis. 

\subsection{Our Multilevel Algorithm}\label{sec:algo}

Suppose \(\{f_1, \ldots, f_m\}\) is a known Markov basis for the model \(\mathcal{M}\). While Lemma \ref{lemma: diaconis_multistep} samples step sizes \(j\) from a discrete distribution with probabilities proportional to inverse factorial weights, our method uses fixed step sizes scaled by level-dependent multipliers to encourage broader exploration of the fiber. To that end, let \(l_1 > l_2 > \cdots > l_L = 1\) denote a sequence of decreasing integer levels. At each level \(l\), we simulate samples \(\{Y_i^{(l)}\}_{i=1}^{N_l}\), where \(N_l\) is the number of samples at level \(l \in \{l_m\}_{m=1}^L\). These samples contribute to the multilevel estimator of the target distribution over chi-square statistics, with higher levels capturing coarse variability and lower levels refining local structure.

To control variance and ensure stable density estimation, we apply a smoothing kernel \(g^{k,\delta}\) to the samples. The kernel function \(g\) is typically a compactly supported, \(r\)-smooth function and the evaluation is performed at a fixed set of grid points \(s_1, \ldots, s_k\). Specifically, for a sample value \(t\), we define
\[
g^{k,\delta}(t) \coloneqq \frac{1}{\delta} \left( g\left( \frac{t - s_1}{\delta} \right), \ldots, g\left( \frac{t - s_k}{\delta} \right) \right).
\]
The final multilevel density estimator is then given by
\begin{equation}\label{eq:ML_estimator}
\widehat{\rho}(s_j) = \frac{1}{N_1} \sum_{i=1}^{N_1} g^{k,\delta}_j\left( Y_i^{(l_1)} \right) +
\sum_{l = l_2}^{l_L} \frac{1}{N_l} \sum_{i=1}^{N_l} \left( g^{k,\delta}_j\left( Y_i^{(l)} \right) - g^{k,\delta}_j\left( Z_i^{(l)} \right) \right)
\end{equation}
for each $j = 1, \ldots, k$, where \(g^{k,\delta}_j(t)\) denotes the \(j\)-th coordinate of \(g^{k,\delta}(t)\), and \(Z_i^{(l)}\) are the coarse-level samples associated with \(Y_i^{(l)}\).

A crucial condition for the validity of the multilevel correction is that the pairs \((Y_i^{(l)}, Z_i^{(l)})\) satisfy the coupling
\[
(Y_i^{(l)}, Z_i^{(l)}) \overset{d}{=} (Y^{(l)}, Y^{(l-1)}),
\]
ensuring unbiasedness of the estimator. That is, the marginal distributions of \(Y^{(l)}\) and \(Y^{(l-1)}\) must be preserved, and their joint distribution must match across adjacent levels. In practice, we approximate this condition by sampling \(Z_i^{(l)}\) uniformly at random from the set of samples obtained at level \(l-1\). While this approach preserves the correct marginals, it does not enforce exact joint dependence across levels. Nevertheless, it yields an unbiased estimator, and more sophisticated coupling strategies could be employed in future work to reduce variance.

Finally, as a post-processing step, we apply Gaussian smoothing to the estimated density over the $k$ discretized domain points to produce a continuous approximation. We will discuss the effect of this additional smoothing on the approximation quality in the next section.

A full algorithmic implementation of our multilevel estimator is given in Appendix \ref{app:experimental}.
In our implementation, we use the conditional volume test and generate uniform samples from the fiber by accepting all feasible Markov moves. Alternative target distributions on the fiber can be obtained by redefining the acceptance probability of the Markov chain when generating the $N_l$ samples in Algorithm \ref{alg:mlmc_fiber}. 

\subsection{Theoretical Guarantees}\label{sec:theoretical}

The theoretical foundation for employing the multilevel density estimator \eqref{eq:ML_estimator} follows from the analysis of~\cite{gil15_multi_density}. The goal is to approximate the true density \(\rho\) of a real-valued random variable \(Y\) at a fixed set of evaluation points \(s_1, \ldots, s_k\). A smoothing kernel \(g^{k,\delta}\) regularizes each sample, ensuring that the resulting density estimate is sufficiently smooth for analysis. 
Crucially, the approximation quality can be controlled by tuning parameters such as the kernel bandwidth \(\delta\), the number of grid points \(k\), and the sample allocation \(N_\ell\) across levels to balance bias and variance.

Under mild assumptions on the smoothness of \(\rho\), the convergence rates of the level-wise approximations, and the regularity of the kernel \(g\), the MSE satisfies the following upper bound (see Equation 3.6 of \cite{gil15_multi_density}):
\begin{align*}
\text{MSE} \lesssim \ &
k^{-2r} + \delta^{2r} \\
&+ \frac{1}{\delta^2} \cdot \min\left( \delta^{-2\alpha_1} M^{-2\alpha_2 l_L},\ M^{-2\alpha_3 l_L} \right) \\
&+ \frac{\log k}{\delta^2} \cdot \left( 
    \frac{1}{N_{l_L}} 
    + \sum_{\ell = l_1}^{l_L - 1} \frac{ \min\left( \delta^{-\beta_4} M^{-\beta_5 \ell},\ 1 \right) }{N_\ell} 
\right)
\end{align*}
where \(k^{-2r}\) is the bias due to grid interpolation with a kernel of smoothness \(r\), \(\delta^{2r}\) reflects the bias introduced by the smoothing kernel bandwidth \(\delta\), the third term captures discretization error at the finest level \(l_L\), and the final term accounts for the sample variance accumulated across levels.

Here, \(M > 1\) is the refinement factor between levels (e.g., halving the step size or increasing resolution), and \(\alpha_1, \alpha_2, \alpha_3, \beta_4, \beta_5 > 0\) are constants that encode the convergence and variance decay rates of the underlying approximation and estimator. This decomposition highlights the trade-offs involved in choosing the kernel bandwidth \(\delta\), the number of evaluation points \(k\), and the number of samples \(N_\ell\) per level.

Importantly, these theoretical guarantees apply only to the estimator evaluated at the discrete grid points \(s_1, \ldots, s_k\). Any continuous density curve produced through interpolation, e.g., via spline smoothing or post-processing with a Gaussian kernel, introduces additional approximation error not captured by this analysis. However, in practice, if the grid is sufficiently fine and the post-processing bandwidth is small, this error is negligible relative to the multilevel estimator's inherent bias and variance.

\section{Measuring Sample Quality}\label{sec:measuring sample quality}

The problem of comparing probability distributions arises across many areas of modern science and engineering. For example, in quasi-Monte Carlo (QMC) methods, a key task is to ``closely'' approximate a target distribution by a discrete set of sampling nodes \cite{HICKKIRKSOR2025, DicEtal14a}. The quality of such approximations is traditionally measured by discrepancy, often with respect to the uniform distribution on the \(d\)-dimensional unit hypercube; a setting intrinsic to classical QMC methods. 

By contrast, in algebraic statistics, where samples are drawn from fibers defined by algebraic constraints, there is no analog of a discrepancy metric to quantify how well a finite sample represents the underlying space. While MCMC practitioners routinely use convergence diagnostics and appeal to concepts like exchangeability \cite{BesagClifford89, Diaconis2022exchangabilityTables}, these provide only indirect assurance of sample quality. A sample may be exchangeable yet still fail to meaningfully capture the structure of the fiber. Motivated by this gap, we consider metrics for comparing discrete distributions, or for evaluating how thoroughly a single sample explores its fiber.

\subsection{Maximum Mean Discrepancy}

The Maximum Mean Discrepancy (MMD) \cite{MMD2012} has been used over the past several decades as a non-parametric approach to determine if two distributions $P$ and $Q$ are different based on samples drawn from them. The MMD quantifies the difference between distributions using functions in a reproducing kernel Hilbert Space (RKHS). Specifically, it measures the discrepancy between two distributions by computing the difference between their mean embeddings in an RKHS. Given a positive definite kernel function \( k: \mathbb{R}^d \times \mathbb{R}^d \to \mathbb{R} \), the MMD is defined as
\begin{equation}
\text{MMD}_k(P, Q) = \sup_{\|f\|_{\mathcal{H}} \leq 1} \Big| \mathbb{E}_{X \sim P}[f(X)] - \mathbb{E}_{Y \sim Q}[f(Y)] \Big|
\end{equation}
where \( \mathcal{H} \) is the RKHS associated with the kernel \( k \). For a probability distribution \( P \), its mean embedding in an RKHS \( \mathcal{H} \) is given by $\mu_P = \mathbb{E}_{X \sim P}[k(X, \cdot)]$ and similarly, for \( Q \), we define \( \mu_Q \) as $\mu_Q = \mathbb{E}_{Y \sim Q}[k(Y, \cdot)]$.
The squared MMD can then be expressed as
\begin{equation}
\text{MMD}^2_k(P, Q) = \| \mu_P - \mu_Q \|^2_{\mathcal{H}}.
\end{equation}
Now, given two independent samples \( X_1, \dots, X_m \sim P \) and \( Y_1, \dots, Y_n \sim Q \), an empirical estimate of \( \text{MMD}^2 \) is
\begin{equation}\label{eq:MMD_est}
\widehat{\text{MMD}}^2_k(P, Q) = \frac{1}{m^2} \sum_{i, j} k(X_i, X_j) + \frac{1}{n^2} \sum_{i, j} k(Y_i, Y_j) - \frac{2}{mn} \sum_{i, j} k(X_i, Y_j).
\end{equation}

The base kernel is often chosen as the distance-induced kernel given by $k(x,x') = ||x||_2+||x'||_2-||x-x'||_2$ for $x,x'\in \mathbb{R}$. In this setting, the MMD reduces to the well known energy-distance \cite{MMDtoenergy2013}.

In this work, we use the MMD to compare samples generated by different procedures against a large reference sample obtained from extensive MCMC simulation. A lower MMD value between a candidate sample and the reference set indicates that the candidate sample more accurately captures the structure of the target distribution and achieves better exploration of the space. Beyond the present study, we hope that the use of MMD as a quantitative sample quality metric will be adopted more widely in algebraic statistics, where there remains a need for principled tools to evaluate how well discrete structures are approximated.

\subsection{Fiber Coverage Score}\label{sec:FCS}

The MMD metric is valuable when one has confidence that the large MCMC reference sample accurately represents the structure of the fiber. However, as emphasized throughout this work, even well-mixed MCMC samples can fail to capture the full geometry of large, high-dimensional fibers. As an alternative, we introduce a \textit{Fiber Coverage Score (FCS)}, designed to assess the quality of a single sample in terms of its exploratory reach within the fiber. This metric is based on a relatively simple idea of partitioning of the polytope $\mathcal{P}_A(b):=\{x\in\mathbb R^n:  Ax=b,  A\in\mathbb Z^{d\times n},b\in\mathbb Z^d\}$ into Voronoi cells, as described below.

\begin{figure}[t]
\centering

\begin{subfigure}{0.4\textwidth}
\centering
\begin{tikzpicture}[scale=3]
\fill[white] (0,0) rectangle (1,1);
\draw[thick] (0,0) rectangle (1,1);

\coordinate (P1) at (0.5,0.25);
\coordinate (P2) at (0.8,0.5);
\coordinate (P3) at (0.4,0.6);

\coordinate (C) at (0.5,0.5);

\foreach \p in {P1,P2,P3}
    \fill (\p) circle (0.8pt);

\begin{scope}
\clip (0,0) rectangle (1,1);

\draw (0,0) -- (C) -- (1,0) -- cycle; 
\draw (1,0) -- (C) -- (1,1) -- cycle; 
\draw (1,1) -- (C) -- (0,0) -- cycle; 

\end{scope}
\end{tikzpicture}
\caption{$K=3$ Voronoi cells.}
\end{subfigure}
\hspace{0.5cm}
\begin{subfigure}{0.4\textwidth}
\centering
\begin{tikzpicture}[scale=3]
\fill[white] (0,0) rectangle (1,1);
\draw[thick] (0,0) rectangle (1,1);

\coordinate (P1) at (0.25,0.45);
\coordinate (P2) at (0.75,0.47);
\coordinate (P3) at (0.5,0.8);
\coordinate (P4) at (0.5,0.4);
\coordinate (P5) at (0.5,0.1);

\foreach \p in {P1,P2,P3,P4,P5}
    \fill (\p) circle (0.8pt);

\coordinate (V12) at (0.5,0.275);
\coordinate (V13) at (0.375,0.55);
\coordinate (V23) at (0.625,0.525);

\begin{scope}
\clip (0,0) rectangle (1,1);

\draw (0,0) -- (V12) -- (V13) -- (0,1) -- cycle;
\draw (1,0) -- (1,1) -- (V23) -- (V12) -- cycle;
\draw (V13) -- (V23) -- (1,1) -- (0,1) -- cycle;

\end{scope}
\end{tikzpicture}
\caption{$K=5$ Voronoi cells.}
\end{subfigure}
\caption{Increasing the number of Voronoi cells for testing robustness of the sample from polytope $\mathcal{P}_A$.}
\label{fig:voronoi}
\end{figure}
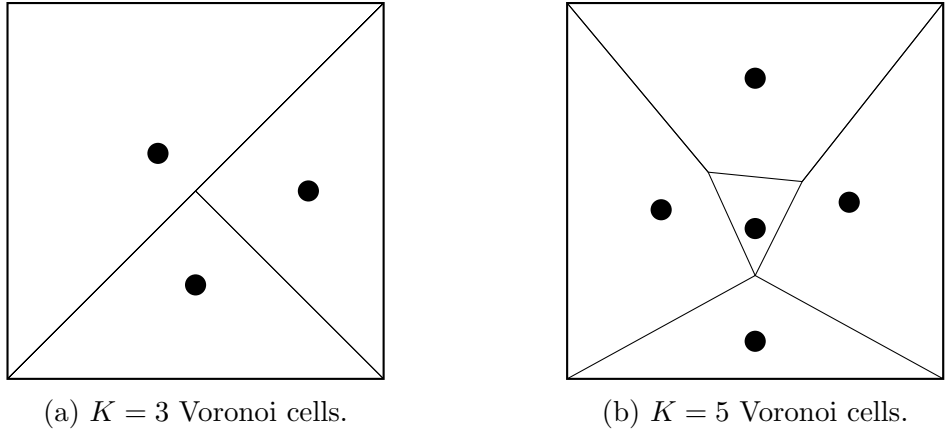

The first step in computing our metric is to generate candidate Voronoi cell centers inside the polytope using a no-drift Brownian motion sampler. The Brownian motion is defined by
\[
Z(t) = Z(0) + \mathcal{B} \sigma B(t),
\]
where \(Z(0) = z_0\) is an initial feasible integer point in the fiber, \(\mathcal{B}\) is a matrix whose columns form a lattice basis, \(\sigma\) is a scale parameter, and \(B(t)\) is a standard Brownian motion (see \cite{Mörters_Peres_2010}). The sampler constructs real-valued combinations of lattice basis vectors using continuous Brownian paths, thereby generating points that lie within the polytope \(\mathcal{P}_A(b)\). Points with negative entries (i.e., \(Z(t) < 0\) component-wise) are rejected and resampled. The resulting \(Z(t)\) are real-valued vectors obeying the constraint \(AZ(t) = b\), and serve as the centers of Voronoi cells. These are visualized as black dots in Figure~\ref{fig:voronoi}. Once we obtain a set of centers \(Z(t)\), we construct Voronoi cells around a selected subset of them and measure sample quality via the following volume-based metric.

\begin{definition}[Fiber Coverage Score]
Let \(\mathbb{R}^d\) be equipped with the Euclidean distance \(d(\cdot, \cdot)\). Let \(\{P_k\}_{k \in K}\) be a set of Voronoi center points and define Voronoi cells
\[
V_k = \left\{ x \in \mathbb{R}^d \,\middle|\, d(x, P_k) \leq d(x, P_j) \text{ for all } j \neq k \right\}.
\]
Given a lattice sample $\{X_i\}_{i=1}^n \subset \fiber{A}(b)$, define the coverage score as
\[
H_K := \frac{1}{K} \sum_{k=1}^{K} \mathbbm{1}_{\{X_i\}_{i=1}^n \cap V_k \neq \emptyset} \in [0, 1].
\]
\end{definition}

The score \(H_K\) quantifies how well the sample 
$\{X_i\}_{i=1}^n$
covers the polytope: if \(H_K \approx 1\) for large \(K\), then the sample intersects most Voronoi regions, suggesting strong coverage of both the polytope and the fiber. Importantly, the score depends on the choice of \(K\). For small \(K\), even highly localized samples may yield a high score. To robustly assess sample quality, we compute \(H_K\) over a range of increasing \(K\). As \(K\) increases, Voronoi cells shrink, and we expect the score to decrease. A high-quality sample is one for which \(H_K\) remains close to 1 even as \(K \to \infty\). To control \(K\) in practice, we generate a large set of Brownian points \(Z(t)\) and regulate their pairwise separation by enforcing a minimum distance threshold. Starting with a large threshold yields a small number of widely separated centers (low \(K\)); gradually reducing the threshold increases \(K\) by allowing denser configurations of Voronoi centers.

Although this metric is designed to evaluate a single sampler, it naturally extends to comparison across samplers. For a fixed fiber and common Voronoi centers \(\{Z(t_0), \dots, Z(t_n)\}\), we compare samplers by computing and averaging \(H_K\) over multiple runs. A sampler is deemed better if it consistently yields higher coverage scores, indicating broader exploration of the fiber space.

\section{Simulations}\label{sec:simulations}

In the experiments that follow (excluding Experiment 2 in Section \ref{sec:multilevel_sims}), we fix the polytope \(\mathcal{P}_A(b)\) defined by the system \(Ax = b\), with fiber \(\fiber{A}(b)\) drawn from a benchmark example in \cite{MB25years}; see Appendix \ref{app:fiber} for full details. The following simulations evaluate the effectiveness of the methods introduced above, with a particular emphasis on the multilevel estimation algorithm presented in Section \ref{sec:algo} (see also Appendix \ref{app:experimental} for implementation details).

Section \ref{sec:multilevel_sims} presents the first implementation of Algorithm \ref{alg:mlmc_fiber}, illustrating the performance of our multilevel sampler. Section \ref{sec:quality_metrics} then evaluates the quality of the resulting samples using the metrics introduced earlier—namely, the Maximum Mean Discrepancy (MMD) and the FCS.

\subsection{Multilevel Sampling} \label{sec:multilevel_sims}

We begin by testing our multilevel sampling procedure on a canonical fiber from the algebraic statistics literature.

\paragraph{Experiment 1.} 

\begin{figure}[t] 
\centering
\includegraphics[scale=0.5]{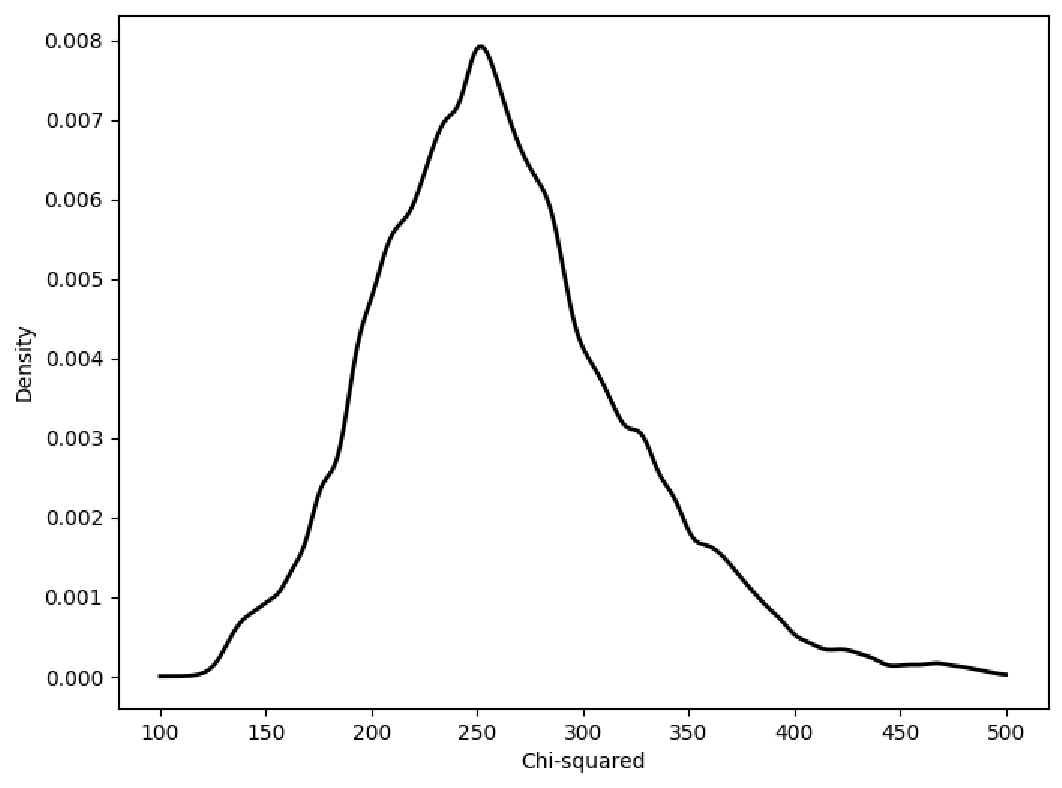}
\caption{Multilevel estimate of the chi-square statistic distribution, obtained via Algorithm 1. The underlying fiber sample is from the uniform distribution.}
\label{fig:multilevel_sampler} 
\end{figure}

We apply the multilevel estimation algorithm to the fiber corresponding to a \(4 \times 4\) contingency table under the model of independence, as described in Appendix \ref{app:fiber}. This example is small but known to pose challenges due to its sparsity. Figure \ref{fig:multilevel_sampler} shows the resulting smoothed density estimate for the chi-square statistic, obtained using the multilevel estimator. Detailed parameter settings and experimental procedures are provided in Appendix \ref{app:experimental}.

This fiber was initially explored in Section \ref{sec:previous_multilevel}, where we discovered issues arising from directly applying the procedure in Lemma \ref{lemma: diaconis_multistep}. 
In order to address these issues, the multilevel sampler operates on both the larger and smaller scales, and obtains multiple Markov chains whose states (fiber points) come from both the local neighborhood around the initial point and the portions of the fiber further away from the initial fiber point. Results stand in stark contrast to the results from Figure \ref{fig: diaconis_multistep_fig} and thus we can already identify advantages of the multi-step approach. The chi-square distribution and the obtained sample cover a greater range of chi-square scores, giving us a good picture of the distribution of the tables in the fiber. In addition, there is no burn-in time in this approach,  and the algorithm obtains the desired distribution in less iterations, as shown in Figure \ref{fig:MMD_MCMCvsMLMC}.

Additionally, we expect that independent runs of the multilevel sampler yield qualitatively similar density estimates. In practice, we observe that repeated executions of our method consistently capture the key features of the target distribution, including the location of the bulk mass and the structure of the tail; we refer to Figure \ref{fig:multiple_runs} in Appendix \ref{app:additional_figs}. This indicates a certain robustness of the sampler in approximating the chi-square statistic distribution over the fiber.

\paragraph{Experiment 2.} Despite the advantages that the multilevel approach yields, the method has an expected shortcoming when it comes to fibers with sufficient statistics with small $L_2$ norm.  In such cases,  all points in the fiber are both sparse and their non-zero integer entries are close to zero. Even with a Markov basis, the sampler won't produce a collection of fiber points for levels that are greater than 1. This is due to the fact that each vector in the Markov basis lives in the null space of constraint matrix 
 and has both negative and positive integer components. Multiplying any of these vectors with a step size (level) $j > 1$ will most likely produce infeasible points, violating the non-negativity constraint. In other words, the move $j\cdot f_I(x)$, will have a negative component greater than any entry in the current point $g(x)$ and thus $g(x) + jf_I(x)$ won't be feasible. 

\begin{figure}[t]
    \centering
\includegraphics[width=0.5\linewidth]{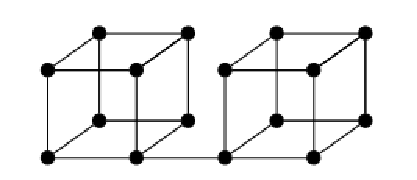}
    \caption{Example of a fiber for which the multilevel approach fails to obtain a sample; see Experiment 2.}
    \label{fig:block_model}
\end{figure}

For a concrete illustration, consider the infamous problem from \cite{Hemmecke2014OnTC}. The fiber $\fiber{A}(b)$  in this case is a integer subset of polytope $\mathcal{P}_{A}$ defined by design matrix 
\begin{align*}A = A_k = 
\begin{pmatrix}
I_k & I_k & \textbf{0} & \textbf{0} & -\textbf{1}_k & \textbf{0}\\
\textbf{0} & \textbf{0} & I_k & I_k & \textbf{0} & -\textbf{1}_k\\
\textbf{0} & \textbf{0} & \textbf{0} & \textbf{0} & \textbf{1} & \textbf{1}
\end{pmatrix} \in \mathbb{Z}^{(2k+1)\times(4k+2)},
\end{align*}
by sufficient statistic $b = \textbf{e}_{2k+1}$ and initial solution $x_0=(\textbf{0}_k, \textbf{1}_k,\textbf{0}_k,\textbf{0}_k,1,0)^T$. The parameter $k$ controls the size of the problem. Visually, the problem looks like Figure \ref{fig:block_model}. The structure of the fiber is such that the fiber graph (pictured) has a single bottleneck edge, which poses a significant challenge for any randomized sampler.  
(The reader may be interested to know that \cite{RUMBA} \emph{does} sample this fiber successfully, after some careful parameter tuning.) 
However, the multilevel sampler encounters a different problem, namely that  the $L_2$ norm of the initial solution $x_0=(\textbf{0}_k, \textbf{1}_k,\textbf{0}_k,\textbf{0}_k,1,0)^T$ is too small, and hence any levels larger than 1 produce unfeasible solutions with negative entries. 

\subsection{Sample Quality}\label{sec:quality_metrics}

We return to sampling from the original fiber detailed in Appendix \ref{app:fiber}.

\begin{figure}[t]
    \centering
    \begin{subfigure}{0.48\linewidth}
        \centering
        \includegraphics[width=\linewidth]{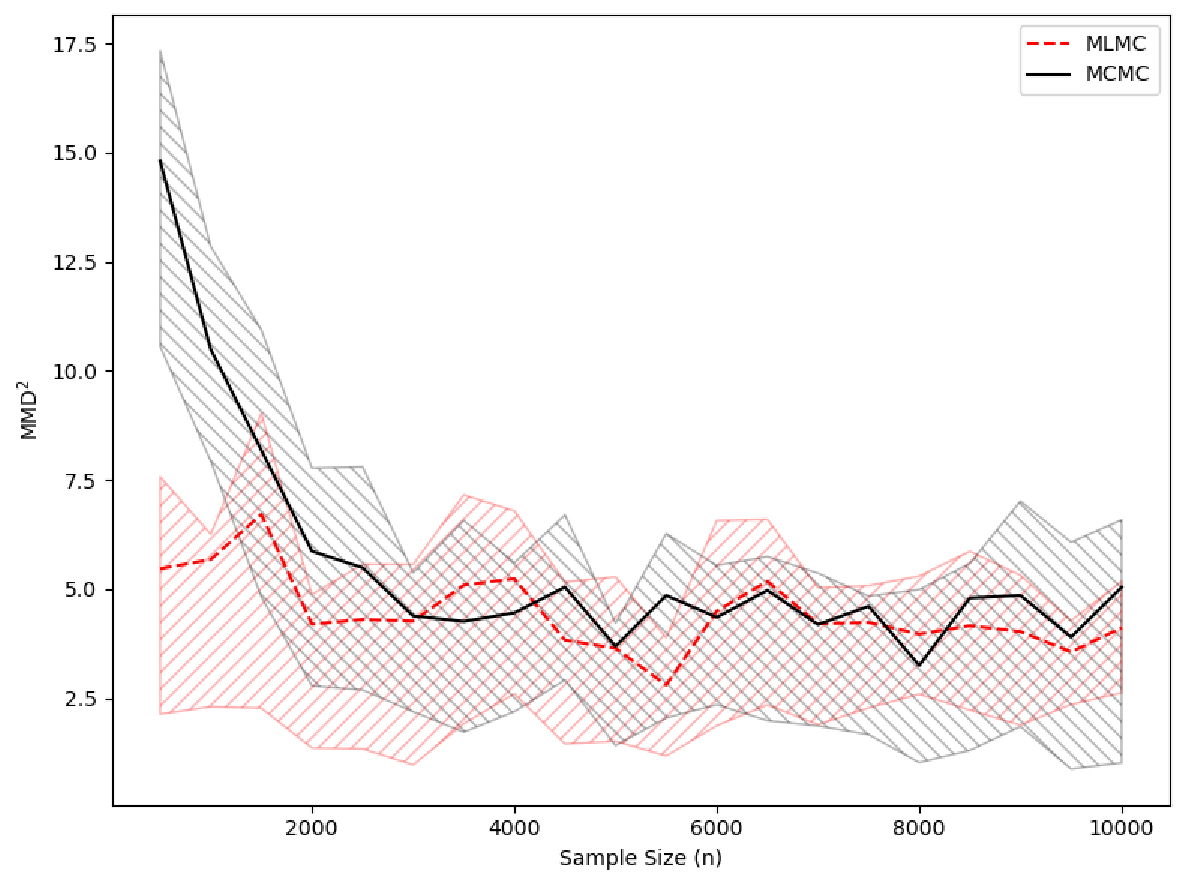}
        \caption{Experiment 3.}
        \label{fig:MMD_MCMCvsMLMC}
    \end{subfigure}\hfill
    \begin{subfigure}{0.48\linewidth}
        \centering
        \includegraphics[width=\linewidth]{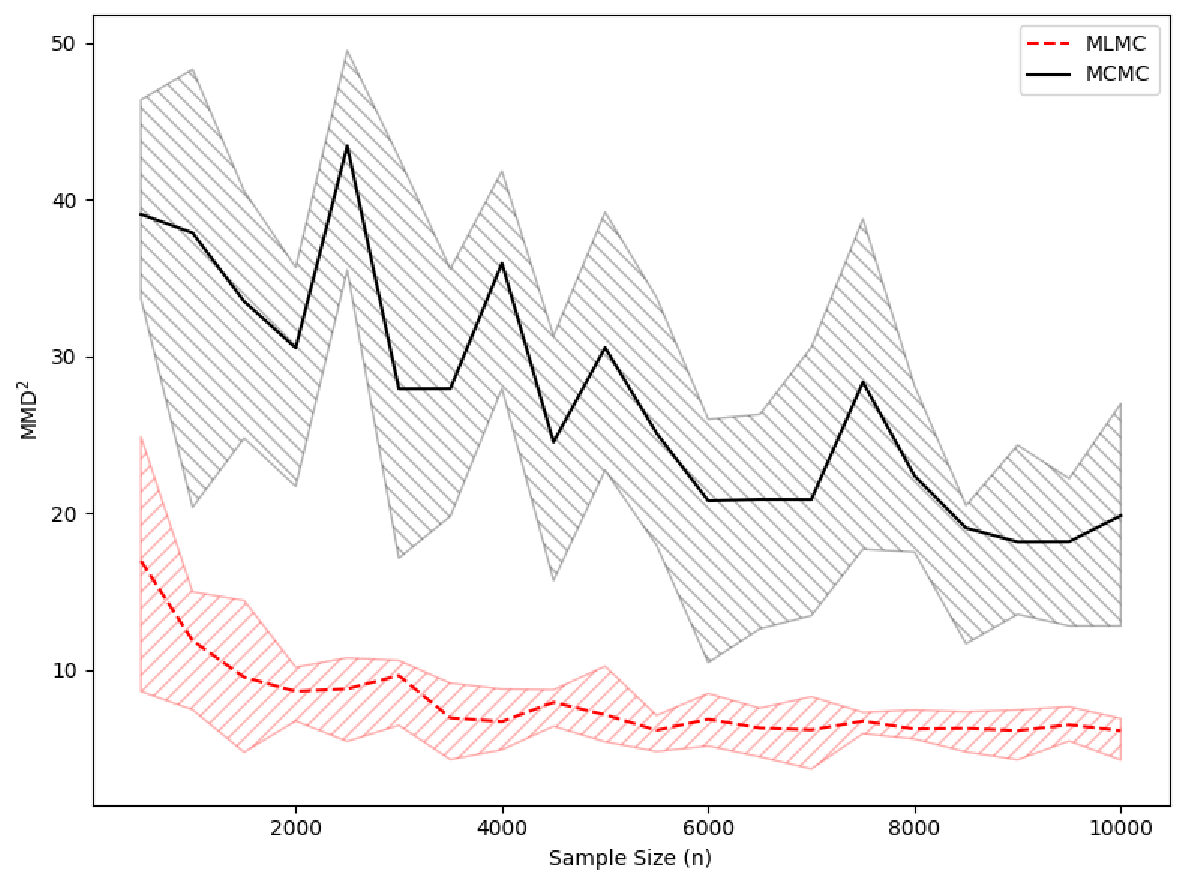}
        \caption{Experiment 4.}
        \label{fig:MMD_other_example}
    \end{subfigure}
    \caption{MMD scores comparing multilevel sampling with regular MCMC. The mean and inter-quartile range of 30 runs is shown.}
    \label{fig:side_by_side}
\end{figure}

\paragraph{Experiment 3.} To test whether multilevel strategies improve sample quality through broader exploration, we compare standard MCMC against our multilevel approach using the Maximum Mean Discrepancy (MMD) statistic defined in Equation \eqref{eq:MMD_est}. As a benchmark, we generate a large reference sample from the fiber of size \(10^5\) using a standard MCMC algorithm with acceptance probability 1 and moves drawn from a Markov basis. This sample serves as a proxy for a representative sample over the fiber. Then, we compute the MMD between this benchmark and samples of size \(n = 500, 1000, 1500, \ldots, 10000\) drawn independently from both the standard MCMC and multilevel methods. The multilevel sampler is implemented using the same experimental setup detailed in Appendix \ref{app:experimental}. Each experiment is repeated over 30 trials, and the resulting mean MMD values with interquartile ranges are reported in Figure \ref{fig:MMD_MCMCvsMLMC}.

Results show that the multilevel approach consistently achieves lower MMD values than standard MCMC for small sample sizes (e.g., \(n = 500\text{--}3000\)), indicating improved approximation of the target distribution with fewer points. As \(n\) increases, the performance gap narrows, and both methods converge to similar MMD values around \(n = 5000\), suggesting comparable approximation quality at larger sample budgets.

\paragraph{Experiment 4.} To demonstrate the effectiveness of the multilevel algorithm on a different fiber and a model more complex than the independence model for two categorical variables, we repeat the procedure of Experiment 3, this time under the no-three-factor interaction model for three categorical variables \cite{FienberBookCategorical}. This model postulates that the conditional relationship between any pair of variables given the third one is the same for all values of the third variable. The exact fiber specification is provided again in Appendix \ref{app:fiber}.

The results in Figure \ref{fig:MMD_other_example} reinforce the conclusions from the previous experiment, and in fact highlight them more strongly. In this case, the fiber is substantially larger, making the advantages of MLMC more pronounced over MCMC. The multilevel exploration rapidly produces samples whose MMD statistic converges to the reference distribution, significantly outpacing the standard MCMC approach.

\paragraph{Experiment 5.} We further compare the multilevel sampler against the approach from Lemma \ref{lemma: diaconis_multistep} and the variant (inverting the probabilities) using the Fiber Coverage Score introduced earlier. This test evaluates the spatial spread of the samples across the fiber. We construct Voronoi cells using a Brownian motion within the polytope \(\mathcal{P}_A\) as described in Section \ref{sec:FCS} and vary the number of cells \(K\) by adjusting the radius of each Voronoi center. As mentioned earlier in the text, the best sampler is the one that maintains a high $H_K$ coverage score even as \(K\) increases.

\begin{figure}[t]
    \centering
    \includegraphics[width=0.7\linewidth]{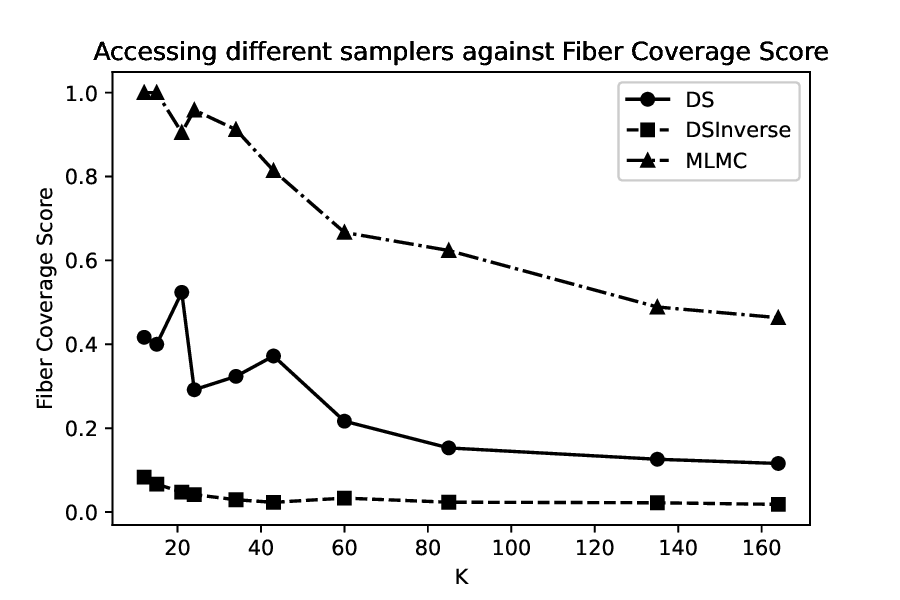}
    \caption{Testing the robustness of three sampling procedures as we increase the number of Voronoi cells (granularity) in the polytope.}
    \label{fig:robust_sample}
\end{figure}

Figure~\ref{fig:robust_sample} displays the behavior of the three sampling strategies. The multilevel sampler outperforms the others, particularly in maintaining high coverage scores at finer granularity. This improved performance is attributed to the coarser levels of the multilevel strategy, which enable more global movement across the fiber, in contrast to the samplers inspired by Lemma \ref{lemma: diaconis_multistep} that tend to remain localized or drift toward the fiber extremes.

\section{Discussion and Future Work}

This work demonstrates how multilevel sampling methods can be adapted effectively for problems in algebraic statistics. Specifically, we show that a multilevel Monte Carlo framework can be used to approximate the distribution of statistics (such as the chi-squared test statistic) over complex discrete spaces defined by algebraic constraints. By leveraging a hierarchy of proposal step sizes derived from a Markov basis and carefully constructing unbiased correction terms across levels, the multilevel procedure achieves improved coverage and sample quality compared to standard MCMC samplers.

Our results suggest that the multilevel methodology is a promising and underutilized tool for sampling and density approximation in algebraic statistics. In particular, we observe superior fiber exploration using the multilevel sampler, as demonstrated both by MMD comparisons to well-mixed MCMC baselines and by our newly introduced Fiber Coverage Score. These results encourage broader adoption of multilevel ideas as a principled sampling strategy in this setting.

While we naturally assume availability of a Markov basis, we would be remiss not to include a discussion on the applicability of multilevel samplers in situations where a Markov basis is not available, as is the case for many problems in practice. We briefly mentioned in the introduction some literature on sampling without the entire basis, e.g. \cite{HAT,DS98}. 
In fact, sampling fibers in the absence of Markov bases can be considered a research area on its own; we single out recent notable work \cite{KahleYoshidaGarciaPuente} that combines Markov chains with sequential importance sampling, as well as incorporating reinforcement learning and Bayesian approaches with incomplete bases \cite{RUMBA} and \cite{RLFiberSampler}. In terms of machine learning, the fiber sampling problem using incomplete bases can be solved, in principle, using a progressive machine learning framework. Toward that end, \cite{IvanPhD2025} introduces the idea of a progression on the coefficients of the linear combination of the lattice basis. Such a construction allows one to start with an easy-to-compute lattice basis and build more complex Markov moves using combinations, the idea being that one can control the size of the coefficients that are used in the linear combination of lattice basis vectors. Depending on the sparsity of the fiber and of the initial solution, a progressive sampler could start with smaller coefficients in a random combination of lattice basis elements, then incrementally increase the coefficients toward the theoretical upper bound. Although this is not a classical machine learning progression, this approach can be thought of as an analogue to the levels in the multilevel sampler presented in our paper. More work is required to fully investigate this direction, but initial simulations show promise that multilevel samplers can be used in this setup. 

Several other avenues remain open for future investigation. First, a key challenge lies in the optimal allocation of computational effort across levels. That is, how should one choose the number of levels $L$, and the number of samples \(N_\ell\) at each level $\ell = 1, 2, \ldots, L$ in order to balance variance decay and cost? While theoretical guidelines exist in the continuous setting \cite{gil15_multi_density}, analogous results in discrete spaces with Markov bases are undeveloped. Following the principles established in the multilevel Monte Carlo literature \cite{Hei01a, giles2008, Gil15a}, one selects $L$ so that the bias at the finest level is of the same order as the sampling error, ensuring that neither source of error dominates. In practice, this can be achieved by performing a short pilot run to estimate how the variance of the level differences decays with $\ell$, and then increasing $L$ until further refinements contribute negligibly to the variance relative to the desired tolerance. Thus, derivaing adaptive or asymptotically optimal allocation rules for algebraic statistics settings is an important next step where fibers may vary dramatically in size and structure: larger or more complex fibers may require deeper hierarchies to adequately control bias, whereas smaller fibers should be handled with fewer levels. Additionally, while our current estimator relies on equal-in-law subsampling between levels to ensure unbiasedness, more sophisticated coupling strategies could improve efficiency further. 
Finally, we again emphasize the importance of developing principled diagnostics for sample quality in discrete spaces. We introduced a Voronoi-based FCS quality metric and demonstrated its use in benchmarking different samplers. Further study of geometric or information-theoretic criteria for fiber exploration could provide new theoretical tools for evaluating and improving discrete sampling algorithms more broadly.

\paragraph{Acknowledgments.} Fred J. Hickernell was instrumental in getting the conversation started with Henry Wynn. Research of first author supported by National Science Foundation (DMS Grant \#2316011). Research of second and third authors partially supported by the Simons Foundation Gift \# 854770 and DOE Grant \#1010629. 

\bibliographystyle{spmpsci}
\bibliography{NMK25.bib,AlgStatAndNtwksAndMB,StatsReferences,henry}

\appendix

\section{Fiber Details}\label{app:fiber}

\paragraph{Experiment 3.} 
Contingency table data arise in many applications, and for a couple of decades applied interest focused on large and high-dimensional tables.  For our purposes, a small but sparse data set suffices to illustrate the difficulties that Markov bases samplers face, and how the multilevel method performs comparatively. Larger experiments are possible, however, Markov bases samplers often take too long and this has been well-documented in the algebraic statistics literature. 

We consider  a synthetic  data set described in \cite[p.371]{DS98}. Note that their discussion is about $10\times10$ tables, but we consider a $4\times4$ sub-example on job-satisfaction, since it already proves our point. Our data is the same $4\times 4$ subset found in \cite[Example 3.13]{MB25years}. The table classifies individuals according to $k=2$ categorical variables: $d_1=4$ levels of salary ranges and $d_2=4$ levels of job satisfaction/dissatisfaction. The table $U$ is thus of size $4\times4$: 

\begin{table}[ht]
\centering
\begin{tabular}{r|rrrr}
  \hline
 & VeryD & LittleD & ModerateS & VeryS \\ 
  \hline
$<$ 15k & 10& 0& 10& 0\\ 
  15-25k & 0& 3& 0& 3\\ 
  25-40k & 0& 0& 2& 40\\ 
  $>$ 40k & 2& 40& 0& 0\\ 
   \hline
\end{tabular}
\end{table}
 and can be flattened into a vector of length $16$:
\\ $u=[10,0,0,2, 0,3,0,40, 10,0,2,0, 0,3,40,0]$.  This is the starting lattice point in our experiments. 

For this example, there are $185,227,230$ tables in the fiber of the model of independence. 
Even if a lattice basis manages to explore the fiber (albeit at a slower rate, as the acceptance probability is usually lower than with Markov bases), the resulting $p$-values are not the same. Specifically, it appears that the lattice basis chain is missing a statistically significant part of the fiber, often not rejecting the model at the nominal level of $0.05$. This is illustrated in \cite[Figures 1 and 2]{MB25years}.

For the model of independence $X_1\perp \!\!\! \perp X_2$, marginals are row and column sums, and thus the equation $Au = t(u)$ is written using the following design matrix: 
\tiny
\[ A = \left[ 
\begin{tabular}{rrrrrrrrrrrrrrrrr}
1 & 1 & 1 & 1 & 0 & 0 & 0 & 0 & 0 & 0 & 0 & 0 & 0 & 0 & 0 & 0 \\ 
0 & 0 & 0 & 0 & 1 & 1 & 1 & 1 & 0 & 0 & 0 & 0 & 0 & 0 & 0 & 0 \\ 
 0 & 0 & 0 & 0 & 0 & 0 & 0 & 0 & 1 & 1 & 1 & 1 & 0 & 0 & 0 & 0 \\ 
 0 & 0 & 0 & 0 & 0 & 0 & 0 & 0 & 0 & 0 & 0 & 0 & 1 & 1 & 1 & 1 \\ 
1 & 0 & 0 & 0 & 1 & 0 & 0 & 0 & 1 & 0 & 0 & 0 & 1 & 0 & 0 & 0 \\ 
 0 & 1 & 0 & 0 & 0 & 1 & 0 & 0 & 0 & 1 & 0 & 0 & 0 & 1 & 0 & 0 \\ 
0 & 0 & 1 & 0 & 0 & 0 & 1 & 0 & 0 & 0 & 1 & 0 & 0 & 0 & 1 & 0 \\ 
0 & 0 & 0 & 1 & 0 & 0 & 0 & 1 & 0 & 0 & 0 & 1 & 0 & 0 & 0 & 1 \\ 
\end{tabular}
\right]\text{\normalsize .}
\] 
\normalsize
One easily checks that 
\[A\cdot 
[10,0,0,2, 0,3,0,40, 10,0,2,0, 0,3,40,0]^T  =  [ 12, 43, 12, 43, 20, 6, 42, 42],\]
 providing the marginals of the the above table. 
 Note that the table\\ $m=[1,-1,0,0, -1,1,0,0, 0,0,0,0, 0,0,0,0]$ satisfies $Am=0$, and is an element of a Markov and lattice bases for $A$. In particular, adding $m$ to the observed table $u$ does not change the margins of the table.  Of course, this move alone cannot be applied to the starting data table, as it would create negative entries, thus stepping outside the set $\fiber A(b)$.  

The job satisfaction table example is a modified version of the following data on job satisfaction from \cite[p.57]{Agresti2002}:
$u = [1,2,1,0, 3,3,6,1,10,10,14,9,6,7,12,11]$. The original job satisfaction data is also used in the documentation page for the R package {\tt algstat} \cite{algstat.R}. It is  not too difficult to analyze, which is why we consider the synthetic data set for which lattice bases samplers already demonstrably do not perform well on average.

\paragraph{Experiment 4.}

Consider the no-three-factor interaction model for three categorical random variables $X,Y,Z$ taking values $\{1,2\}\times\{1,2\}\times\{1,\dots,5\}$. For this model, also often called the model of homogeneous associations in the contingency tables literature, the toric matrix $A$ which computes the table marginals is the following: 

\tiny
\[ A = \left[ 
\begin{tabular}{rrrrrrrrrrrrrrrrrrrr}
1&0&0&0&1&0&0&0&1&0&0&0&1&0&0&0&1&0&0&0\\
0&1&0&0&0&1&0&0&0&1&0&0&0&1&0&0&0&1&0&0\\
0&0&1&0&0&0&1&0&0&0&1&0&0&0&1&0&0&0&1&0\\
0&0&0&1&0&0&0&1&0&0&0&1&0&0&0&1&0&0&0&1\\
1&1&0&0&0&0&0&0&0&0&0&0&0&0&0&0&0&0&0&0\\
0&0&1&1&0&0&0&0&0&0&0&0&0&0&0&0&0&0&0&0\\
0&0&0&0&1&1&0&0&0&0&0&0&0&0&0&0&0&0&0&0\\
0&0&0&0&0&0&1&1&0&0&0&0&0&0&0&0&0&0&0&0\\
0&0&0&0&0&0&0&0&1&1&0&0&0&0&0&0&0&0&0&0\\
0&0&0&0&0&0&0&0&0&0&1&1&0&0&0&0&0&0&0&0\\
0&0&0&0&0&0&0&0&0&0&0&0&1&1&0&0&0&0&0&0\\
0&0&0&0&0&0&0&0&0&0&0&0&0&0&1&1&0&0&0&0\\
0&0&0&0&0&0&0&0&0&0&0&0&0&0&0&0&1&1&0&0\\
0&0&0&0&0&0&0&0&0&0&0&0&0&0&0&0&0&0&1&1\\
1&0&1&0&0&0&0&0&0&0&0&0&0&0&0&0&0&0&0&0\\
0&0&0&0&1&0&1&0&0&0&0&0&0&0&0&0&0&0&0&0\\
0&0&0&0&0&0&0&0&1&0&1&0&0&0&0&0&0&0&0&0\\
0&0&0&0&0&0&0&0&0&0&0&0&1&0&1&0&0&0&0&0\\
0&0&0&0&0&0&0&0&0&0&0&0&0&0&0&0&1&0&1&0\\
0&1&0&1&0&0&0&0&0&0&0&0&0&0&0&0&0&0&0&0\\
0&0&0&0&0&1&0&1&0&0&0&0&0&0&0&0&0&0&0&0\\
0&0&0&0&0&0&0&0&0&1&0&1&0&0&0&0&0&0&0&0\\
0&0&0&0&0&0&0&0&0&0&0&0&0&1&0&1&0&0&0&0\\
0&0&0&0&0&0&0&0&0&0&0&0&0&0&0&0&0&1&0&1
\end{tabular}
\right]\text{\normalsize .}
\] 
\normalsize

For the experiment, we generated $570$ random triples of joint observations of $X,Y,Z$, which can be classified in the following  $2\times2\times5$ contingency table:  
\[
\begin{tabular}{|rr|}
\multicolumn{2}{c}{$Z=1$}\\
\hline
39&27\\23&23\\
\hline
\end{tabular}
\quad
\begin{tabular}{|rr|}
\multicolumn{2}{c}{$Z=2$}\\
\hline
25     & 28\\  31     &  25\\
\hline
\end{tabular}
\quad
\begin{tabular}{|rr|}
\multicolumn{2}{c}{$Z=3$}\\
\hline
 42   &   34\\ 32&      28\\
\hline
\end{tabular}
\quad
\begin{tabular}{|rr|}
\multicolumn{2}{c}{$Z=4$}\\
\hline
 33  &    24\\ 21 &     27\\
\hline
\end{tabular}
\quad
\begin{tabular}{|rr|}
\multicolumn{2}{c}{$Z=5$}\\
\hline
26   &   27\\   23 &     32\\
\hline
\end{tabular}. 
\]
In the random data set, there were $39$ observations $\{X=1,Y=1,Z=1\}$, $27$ observations $\{X=1,Y=2,Z=1\}$, and so on. 
 Flattened into a vector in $\mathbb Z^{20}$, the following is then the starting lattice point for the samplers: 
\[u=[39,   23,   27,   23,   25,   31,   28,   25,   42,    32,    34,    28,    33,    21,    24,   27,    26,    23,    27,    32].\]

The Markov basis for this model consists of $10$ elements and can be found, for example, on the website \cite{MBdatabase}, or can be computed using a computer algebra software. This model, being the smallest non-decomposable model for contingency tables, is notorious for its Markov bases complexity \cite{slimTables}, which is unbounded as the number of states of the variables grows.

\section{Experimental Implementation}\label{app:experimental}

The simulations in Section \ref{sec:multilevel_sims} used a four-level multilevel scheme. At level $\ell$, the number of samples was set to $N_\ell = \lfloor 100{,}000 / 2^{\ell - 1} \rfloor$, and the Markov move size was scaled by a level-dependent multiplier $m_\ell = 5 - \ell$, yielding multipliers $[4, 3, 2, 1]$ from finest to coarsest levels. Proposed moves resulting in negative entries were rejected deterministically.

To estimate the marginal distribution of the chi-squared statistic over the fiber, we applied kernel density estimation at each level using the Epanechnikov kernel with bandwidth $\delta = 1.0$, evaluated on 400 grid points between 100 and 500. The final estimate was smoothed with a Gaussian filter.

Below is the full Algorithm implementing the multilevel estimator contained in Section \ref{sec:algo}.

\begin{algorithm}[htbp]
\caption{Multilevel Density Estimator for Fiber Sampling}
\label{alg:mlmc_fiber}
\begin{algorithmic}
\Require Markov basis $\{f_1, \ldots, f_m\}$ for model $\mathcal{M}$, levels $\{l_1 > l_2 > \cdots > l_L = 1\}$, sample sizes $\{N_{l}\}_{l=1}^{L}$, grid points $\{s_j\}_{j=1}^k$, smoothing kernel $g$, bandwidth $\delta$
\Ensure Smoothed density estimate $\widehat{\rho}(s_j)$ for $j = 1, \ldots, k$

\For{$l \in \{l_1, \ldots, l_L\}$}
    \State Generate $N_l$ samples $\{Y_i^{(l)}\}_{i=1}^{N_l}$ using a random walk over the fiber with fixed step size scaled by $l$
\EndFor

\For{$l \in \{l_2, \ldots, l_L\}$}
    \State For each $i = 1, \ldots, N_l$, set $Z_i^{(l)} \gets$ random draw from $\{Y_i^{(l-1)}\}_{i=1}^{N_{l-1}}$
\EndFor

\State Initialize $\widehat{\rho}(s_j) \gets 0$ for $j = 1, \ldots, k$
\For{$i = 1$ to $N_1$}
    \For{$j = 1$ to $k$}
        \State $\widehat{\rho}(s_j) \mathrel{+}= \frac{1}{N_1} \cdot \frac{1}{\delta} \cdot g\left( \frac{Y_i^{(l_1)} - s_j}{\delta} \right)$
    \EndFor
\EndFor

\For{$l \in \{l_2, \ldots, l_L\}$}
    \For{$i = 1$ to $N_l$}
        \For{$j = 1$ to $k$}
            \State $\widehat{\rho}(s_j) \mathrel{+}= \frac{1}{N_l} \cdot \left[ \frac{1}{\delta} g\left( \frac{Y_i^{(l)} - s_j}{\delta} \right) - \frac{1}{\delta} g\left( \frac{Z_i^{(l)} - s_j}{\delta} \right) \right]$
        \EndFor
    \EndFor
\EndFor

\State Smooth $\widehat{\rho}(s_j)$ with a Gaussian filter across $j=1,\ldots,k$
\State \Return $\{\widehat{\rho}(s_j)\}_{j=1}^k$
\end{algorithmic}
\end{algorithm}

\section{Additional Figures}\label{app:additional_figs}

\begin{figure}[h]
    \centering
    \includegraphics[width=0.9\linewidth]{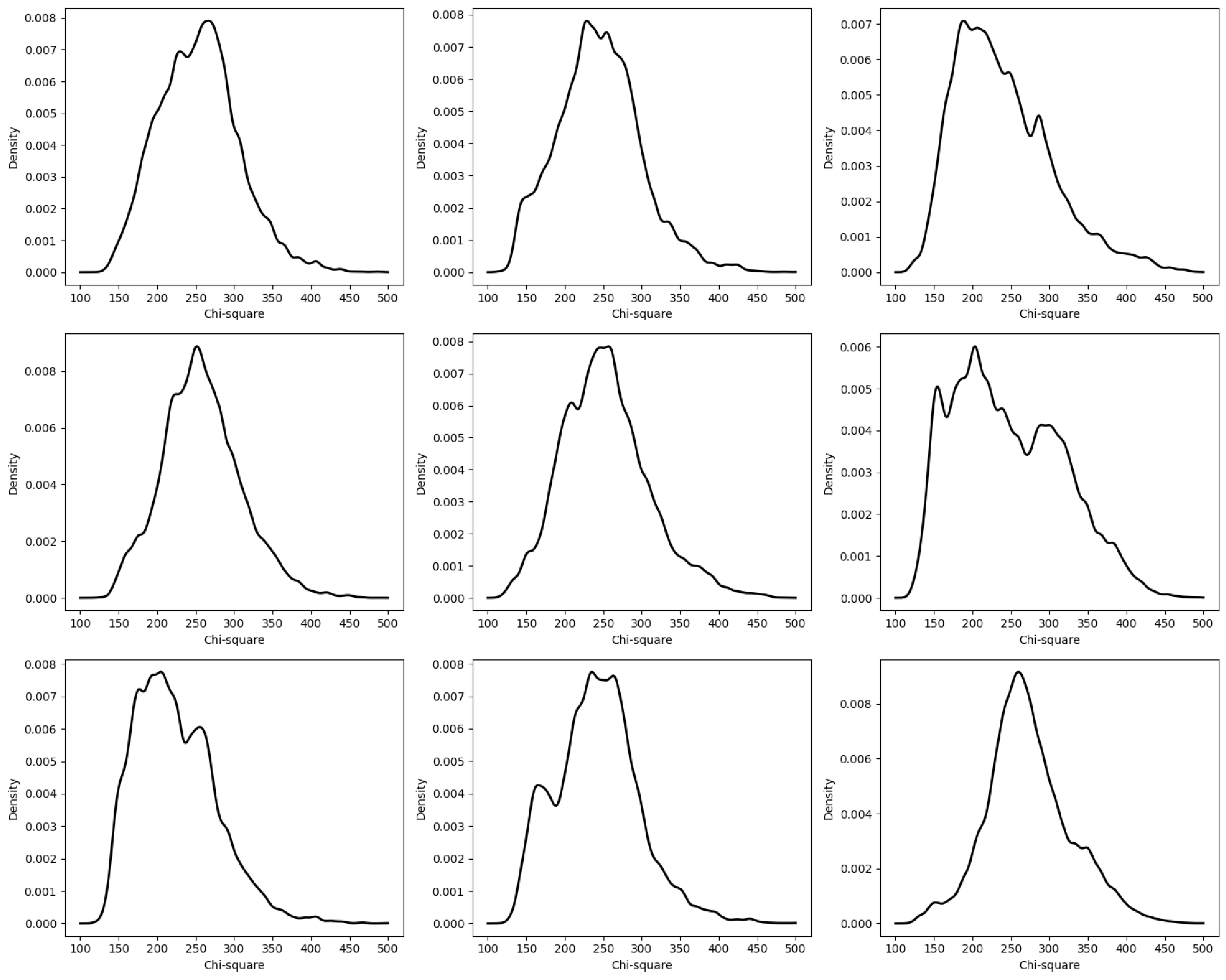}
    \caption{Independent runs of multilevel procedure in Algorithm \ref{alg:mlmc_fiber}}
    \label{fig:multiple_runs}
\end{figure}

Extending from Figure \ref{fig:multilevel_sampler}, we show in Figure \ref{fig:multiple_runs} that multiple independent runs of the multilevel estimator yield consistent density estimates, demonstrating the method’s stability and reproducibility.

\end{document}